# Personality Profiles of Software Engineers and Their Software Quality Preferences

Arif Raza, National University of Sciences & Technology, Pakistan

Luiz Fernando Capretz, The University of Western Ontario, Canada

Zaka-ul-Mustafa, National University of Sciences & Technology, Pakistan

## ABSTRACT

Studies related to human aspects in software engineering (SE) have been performed from different perspectives. These perspectives include the study of human factors in different phases of software life cycle, effect of team performance in software development, how can a personality trait suit a particular task, and about some other miscellaneous issues. This research work aims to establish personality profiles of Pakistani software engineers using the Myers-Briggs Type Indicator (MBTI) instrument. In this survey, we have collected personality profiles of 110 software engineers. Moreover, their preferences of software quality attributes have also been collected. Analysis of the study shows that the most prominent personality type is a combination of introversion, sensing, thinking and judging. Investigative results indicate that most of the software engineers consider usability and functionality as the most important software quality attributes.

## INTRODUCTION

Human factors in software engineering have different dimensions. Studies in this domain have been performed from different perspectives. Looking back two or three decades, software engineers did not have to do much social interaction in their jobs; however the situation has changed now. Human resources people do not consider knowledge in applied computing or software engineering sufficient enough to hire software professionals. They expect these candidates to have the ability to learn, capability to work in teams, oral and written communications skills, and orientation to health and wellness. In short, adaptability, communication, and stress management are seen as key talents for software professionals.

Developing a high quality and efficient software has always been a high priority in the software industry. The International Organization for Standardization and the International Electro-technical Commission, ISO/IEC 9126-1 (2001) classifies software quality attributes into six categories: functionality, reliability, usability, efficiency, maintainability and portability.

MBTI has been used in studies to illustrate personality profiles of general population as well as of software professionals (Bishop-Clark & Wheeler, 1994; Capretz, 2003; Da Cunha & Greathead, 2007; Miller & Zhichao, 2004). However, not many references are available related to South-Asian software professionals. To acquire and study personality profiles of software engineers is vital to develop software industry in this region.

The main objective of this work is to study personality profiles of Pakistani software engineers and their preference of software quality attributes. In this survey, we have collected personality profiles of 110 software engineers using MBTI questionnaire. Their software quality preferences have also been gathered and analyzed.

## THE MYERS-BRIGGS TYPE INDICATOR (MBTI)

The Myers-Briggs Type Indicator (MBTI) (Meyers, McCaulley, Quenk, & Hammer, 1998) is mainly used for the classification of personality types. It has also been used to comprehend individual learning styles and preferences in inspiration. The MBTI has four dimensions of preferences, which describe a specific personality. Within each dimension, there are two opposite pairs: Extroversion - Introversion, Sensing - Intuition, Feeling - Thinking, and Perceiving - Judging. Consequently, sixteen typical personality types are defined by using combination of these four distinct types, each denoted by four letters. Although, it is possible for an individual to use all eight preferences in each of the four pairs, generally every person has one dimension that is more dominant in his/her personality than its complementary part. The scales are briefly described in Appendix-B:

MBTI defines sixteen types to describe people's personalities, temperaments and approaches towards general issues of life. For example, if an individual is found to be the ISTP type, it means that the individual prefers Introversion, Sensing, Thinking, and Perceiving. This also signifies compatibility of personality types with a specific job and how one makes decisions in different situations. Although, these categories may sustain enhanced performance in some conditions, no category can be considered better than the other.

## LITERATURE REVIEW

Many empirical studies have been conducted to investigate the relationship between MBTI and software engineers. Most of them indicated ISTJ as the most frequent personality type. In a study which involved 58 software professionals, Bush and Schkade (1985) found ISTJ (25%) as the most common personality type, followed by INTJ (16%), and ENTP (9%). Results of Buie (1988) indicated ISTJ (19%), INTP (15%) and INTJ (13%) based on the data collected from 47 scientific computer professionals. ESFJ (0%), ISFP (0%) and ENTP (0%) were however under-represented. The most frequent types in the sample of 37 systems analysts as studied by Smith (1989) were ISTJ (35%) and ESTJ (30%). Lyons (1985) survey of 1229 software professionals from over 100 companies found ISTJ (23%) to be the most common type, INTJ (15%) to be the second, closely followed by INTP (12%).

Capretz (2003) investigated the profile of a group of 100 productive and motivated software engineers. After assessing their personality types through MBTI (Form G), ISTJ was found to be the largest single type among the participants in his study. Considering the dominance of introverts in the software field, he concluded, *"the majority of software engineers (ISTJ) are technically oriented and prefers working with facts and reason rather than with people."*

Varona et.al. (2011) established the personality profile of Cuban software engineers using MBTI. Analysis of their study indicated ESTJ as the most prominent personality type. The authors maintained adaptability, communication and stress management as main expertise for software engineers. They foresaw more widely distribution of extroverts than introverts in the software industry in future.

Past researches have been carried out to study effects of personalities in different dimensions. To find out whether the pair programming reduces the time required to solve a computer programming job. Arisholm, Gallis, Dyba, & Sjoberc (2007) conducted a study with professional Java programmers as subjects. Based on 29 international consultant companies in Europe, in the study 295 junior, intermediate and senior professionals used Java tools on two alternate systems with a varying degree of complexity. In the context of system complexity and programmers' expertise, effects of pair programming were evaluated. The results were based on the correctness of solution and maintenance tasks on some code written in Java programming language.

Karn and Cowling (2006) highlighted the effects of different personality types using MBTI on the working of software engineering teams. The study described how ethnographic methods could be used to study SE teams, to understand the role of human factors in a SE project. The results of the study indicated how cohesive teams could be constructed with the help of the knowledge about psychological types of team members. The study also highlighted *"the conceptual orientation of software engineers"* while going through different phases of a project. Bradley and Hebert (1997) and Acuna et. al. (2009) in their studies also agreed that team performance was affected by personality types.

Software development life cycle comprises of different phases such as requirements, design, implementation, testing and maintenance. Realizing the importance of all personality types for software engineering, Capretz and Ahmed (2010) stated that every type could make a contribution towards problem solving. However, their study concluded that specific personality dimension might contribute more to one particular phase but less to another phase of the software life cycle. They did not rule out the possibility of affecting certain stages of software engineering in different ways, by certain personality profiles.

In their study Liao and Huang (2009) maintain that there exists a positive relationship between community adaptability and perceived ease of use and usefulness of e-learning. The study highlights *"perceived ease of use and attitude"* as the two factors which affect students' behaviour towards e-learning.

## RESEARCH METHODOLOGY

In this study we surveyed 110 Pakistani software engineers that included SE students and professors of National University of Sciences and Technology, Islamabad, Pakistan. A short version of the MBTI form (form G) was provided to identify their personality types. They were invited to take the MBTI measure at the university campus. The criteria to select the students to take part in this survey included their interest in software development projects as well such as in taking MBTI test. Grade Point Averages (GPAs) of the students, however, were not taken into account. There were 64 final year under graduate (51 males, 13 females) students, 28 post graduate (18 males and 10 females) students and 18 professors (15 males, 3 females). The students' age range was between 21 and 23, whereas professors' age range was between 28 and 45 years old.

Overall, the objective of this study is to investigate the answer to the following question:

*"Which quality attributes software engineers consider important for their projects?"*

In order to empirically investigate the research question, following hypotheses are derived. The independent variables are selected from the six categories of software quality attributes as classified by ISO/IEC 9126-1 (2001). Definitions of the attributes are given in Appendix-A.

**H1:** Software engineers consider functionality as an important quality attribute for their software projects.

**H2:** Software engineers consider usability as a significant quality attribute for their software projects.

**H3:** Software engineers consider efficiency as an important quality attribute for their software projects.

**H4:** Software engineers consider maintenance as a vital quality attribute for their software projects.

**H5:** Software engineers consider reliability as an imperative quality attribute for their software projects.

**H6:** Software engineers consider portability as an essential quality attribute for their software projects.

The participants were asked to indicate their preference with respect to six software quality attributes of ISO 9126-1 (2001) i.e. functionality, reliability, usability, efficiency, maintainability and portability. They were asked, *"Which one software quality attribute do you personally consider most important in your software project?"*

**Data Analysis**

The personality type distribution of the participants is summarized in Table 1 and Table 2 below. It can be observed that among our respondents, introverts (58%) are more than the extroverts (42%). Similarly, sensing (59%) dominate over intuitive (41%) and thinking (60%) over feeling (40%), whereas judging (48%) and perceiving (52%) are close. All percentage values have been rounded to the nearest decimal.

When analyzing the results with reference to the individual poles it is clear that 'Ts' and 'Ss' (with 60% and 59% respectively) are over-represented, whereas 'Fs' and 'Ns' (with 40% and 41% each) are under-represented.

Table 1 Software Engineers personality types distribution in each dimension

| I | E |
|---|---|
| 58 % | 42 % |
| **N** | **S** |
| 41% | 59 % |
| **T** | **F** |
| 60 % | 40 % |
| **J** | **P** |
| 48 % | 52 % |

Table 2 shows that ISTJ personality type is the most represented type (14%), followed by ENTP (11%), and ESTJ and ISFJ (9% each). These numbers represent almost half of the sample. The least represented combinations is ENTJ (1%) followed by ENFJ (2%) and INFJ (3 %).

Table 2. The MBTI Types and their Distribution among the Pakistani Software Engineers

| ISTJ | ISFJ | INFJ | INTJ |
|---|---|---|---|
| 14 % | 9 % | 3 % | 7 % |
| **ISTP** | **ISFP** | **INFP** | **INTP** |
| 6 % | 7 % | 4 % | 8 % |

| ESTP | ESFP | ENFP | ENTP |
|------|------|------|------|
| 4 %  | 6 %  | 5 %  | 11 % |
| **ESTJ** | **ESFJ** | **ENFJ** | **ENTJ** |
| 9 %  | 4 %  | 2 %  | 1 %  |

Software quality attribute preference data is shown in Table 3. Functionality is the most popular quality attribute preferred by 32%, closely followed by usability, which is considered most important by 31% of the participants. These are followed by reliability (19%) and efficiency (16%). However portability (2%) and maintainability (3%) are found at the bottom of the priority list of the participated software engineers.

Table 3. The Quality Attribute Preference of Pakistani Software Engineers

| Quality Attribute | Frequency (%) |
|---|---|
| Functionality | 32 |
| Usability | 31 |
| Efficiency | 16 |
| Maintainability | 3 |
| Reliability, | 16 |
| Portability | 2 |

Individual indicator distribution is presented in Table-4. According to this survey usability and functionality are the two quality attributes preferred by all the respondents.

Table 4: Individual Indicator Distribution

| Type | Quantity | Functionality | | Usability | | Efficiency | | Maintenance | | Reliability | | Portability | |
|---|---|---|---|---|---|---|---|---|---|---|---|---|---|
| I | 64 | 27% | 17 | 31% | 20 | 19% | 12 | 3% | 2 | 17% | 11 | 3% | 2 |
| E | 46 | 39% | 18 | 31% | 14 | 13% | 6 | 2% | 1 | 15% | 7 | 0% | 0 |
| S | 65 | 29% | 19 | 31% | 20 | 15% | 10 | 0% | 0 | 23% | 15 | 2% | 1 |
| N | 45 | 35% | 16 | 31% | 14 | 18% | 8 | 7% | 3 | 7% | 3 | 2% | 1 |
| T | 66 | 36% | 24 | 29% | 19 | 14% | 9 | 4% | 3 | 14% | 9 | 3% | 2 |
| F | 44 | 25% | 11 | 34% | 15 | 23% | 10 | 0% | 0 | 18% | 8 | 0% | 0 |
| J | 53 | 19% | 10 | 38% | 20 | 21% | 11 | 0% | 0 | 19% | 10 | 3% | 2 |

| P | 57 | 44% | 25 | 25% | 14 | 12% | 7 | 5% | 3 | 14% | 8 | 0% | 0 |

## Reliability Analysis of Measuring Instrument

The reliability is considered an integral feature of an empirical study. It indicates the reproducibility of a measurement. The reliability of the measurement scales of the six quality attributes is evaluated using an internal-consistency analysis by means of coefficient alpha (Cronbach, 1951). In the analysis, the range of coefficient alpha is from 0.55 to 0.65 as presented in Table -5. van de Ven and Ferry (1980) consider a reliability coefficient of 0.55 or higher as satisfactory, and Osterhof (2001) state that 0.60 or higher is acceptable. Therefore, it is concluded that the variable items developed for this empirical investigation are reliable.

Table 5: Coefficient Alpha values of variables

| Quality Attributes | Coefficient α |
|---|---|
| Functionality | 0.65 |
| Usability | 0.60 |
| Efficiency | 0.55 |
| Maintenance | 0.61 |
| Reliability | 0.64 |
| Portability | 0.61 |

## Hypotheses Testing

To test the hypotheses H1-H6, parametric statistics were used to examine the Pearson correlation coefficient between software engineers and the software quality attributes. The results of the statistical calculations for the Pearson correlation coefficient are displayed in Table 6. *"In statistical hypothesis testing, the p-value is the probability of obtaining a test statistic. The lower the p-value, the less likely the result is if the null hypothesis is true, and consequently the more "significant" the result is, in the sense of statistical significance"* (Wikipedia).

The Pearson correlation coefficient between software engineers and functionality was found to be positive (0.552) at $P > 0.05$, and hence did not justify the hypothesis H1. A Pearson correlation coefficient of 0.750 was observed at P < 0.05 between software engineers and usability and hence was found to be significant.

Table 6: Hypotheses testing using parametric correlation coefficients

| Hypothesis | Software Quality Attribute | Pearson Correlation Coefficient |
|---|---|---|
| H1 | Functionality | 0.552<br>P = 0.156** |
| H2 | Usability | 0.750<br>P = 0.032* |
| H3 | Efficiency | 0.414<br>P = 0.308** |
| H4 | Maintenance | 0.194<br>P = 0.644** |
| H5 | Reliability | 0.724<br>P = 0.042* |
| H6 | Portability | 0.594<br>P = 0.120** |

* Significant at P < 0.05. ** Insignificant at P > 0.05.

The hypothesis H3 and H4 were rejected based on the Pearson correlation coefficient (0.414) at P > 0.05, between the software engineers and efficiency, and the correlation coefficient of 0.194 at P >0.05 between the software engineers and maintenance, respectively. However, hypothesis H5 was found to be significant after analyzing the Pearson correlation coefficient of 0.724 at P = 0.042 between usability the software engineers and reliability. Finally, the hypothesis H6 was rejected based on the Pearson correlation coefficient (0.594) at P > 0.05, between the software engineers and portability. Hence, as observed and reported above, hypotheses H2 and H5 were found to be statistically significant and were accepted whereas H1, H3, H4 and H6 were not supported and therefore got rejected.

**DISCUSSION – THEORETICAL AND PRACTICAL IMPLICATIONS**

The findings of this study have several theoretical and practical implications. On the theoretical front, the study highlights the role of individual differences and personality profiles of Pakistani Software Engineers community. According to the above results, introverts have higher preference for usability, whereas extroverts consider functionality as the most important quality attribute for their software. Sensings and intuitives did not show much difference as far as their preference for top two quality attributes is concerned. In the case of the thinking-feeling poles, higher value of feeling's preference for usability is notable, whereas thinkers show a remarkable preference for functionality. Judgers have got a higher value for usability too, as compared to perceivers who prefer functionality. It is also to be noted that portability and maintainability are the two least preferred quality attributes for all the types, in general.

One practical implication of this study is that the human resource people of different organizations could consider these traits when employing software engineers. In this way, they could hire people with personality profiles that better match the overall policy of the organization.

**Limitations of the Study and Threats to External Validity**

Singer and Vinson (2002) consider surveys, experiments, metrics, case studies, and field studies as examples of empirical methods used to investigate both software engineering processes and products. Surveys are subject to certain limitations which is the case of this study too. Wohlin, Runeson, Host, Ohlsson, Regnell, & Wesslen (2000) maintain that threats to external validity are conditions that limit researchers' abilities to generalize the results of their experiments to industrial practices. Our survey population is software professionals; the sample is thus biased towards those personality types that tend to be present in that population. Specific measures were however taken to support external validity, for example, a random sampling technique was used to select the respondent from the population in order to conduct experiments.

Faden, Beauchamp, & King (1986) emphasize that the increased popularity of empirical methodology in software engineering has raised apprehensions on the ethical issues. We followed the recommended ethical principles to ensure that the empirical investigation conducted and reported here would not violate any form of recommended experimental ethics. Our subjects took part in the survey voluntarily and they were not offered any compensation in any form.

Another limitation of this study is its relatively small sample size in terms of number of respondents. Although the proposed approach has some potential to threaten external validity, we followed appropriate research procedures by conducting and reporting tests to improve the reliability of the study, and certain measures were also taken to ensure the external validity.

**CONCLUSION**

Most studies relating to MBTI distribution among engineers exhibit that ISTJ, INTJ and ESTJ are over-represented personality types, whereas ENFJ and INFJ are under-represented personalities (Capretz, 2003; Miller & Zhichao, 2004). The results of our study also follow the same trend with ISTJ personality type as the most common personality type. However, in contrast with the trends of existing studies, ENTP, ESTJ and ISFJ have also been over represented.

The results of our study indicate over-representation of Introverts and ISTJs. The preference of introverts for usability as a dominant software quality attribute indicates software engineers' trend towards user centred software designs.

We foresee improved reliability and better variance with increase in the number of sample size. We are currently in contact with local software industry to conduct similar surveys.

**Appendix-A**

The International Organization for Standardization and the International Electro-technical Commission, ISO/IEC 9126-1 (2001) classifies software quality attributes into six categories as defined below [1]:

**Functionality** – "*A set of attributes that bear on the existence of a set of functions and their specified properties. The functions are those that satisfy stated or implied needs.*"

**Reliability** – "*A set of attributes that bear on the capability of software to maintain its level of performance under stated conditions for a stated period of time.*"

**Usability** – "*A set of attributes that bear on the effort needed for use, and on the individual assessment of such use, by a stated or implied set of users.*"

**Efficiency** – "*A set of attributes that bear on the relationship between the level of performance of the software and the amount of resources used, under stated conditions.*"

**Maintainability** – "*A set of attributes that bear on the effort needed to make specified modifications.*"

**Portability** – "*A set of attributes that bear on the ability of software to be transferred from one environment to another.*"

**Appendix-B**

**The Myers-Briggs Type Indicator (MBTI) Scales**

Extroversion (E) – Introversion (I): Extroverts prefer to communicate with other people by focusing on outer world of people and things, whereas Introverts choose to work independently by focusing on inner world of ideas and emotions.

Sensing (S) - Intuition (N): This dimension is about the way people gain information. Sensing people trust on their experience and tend to focus on facts they can count on, while Intuitive individuals are more focused on their creativity, insight and new potential of events.

Thinking (T) - Feeling (F): The third dimension is about the way people take decisions in life. Thinking individuals are cool headed, prefer clearly defined tasks and have a logical and analytical reasoning to make decisions, whereas Feeling people are warm hearted, consider harmonious working relationship important and have a sensitive approach.

Judging (J) - Perceiving (P): Judging type like to follow a schedule, prefer to have things settled and do not like too much spontaneity whereas Perceiving type prefer to keep their options open to alteration, prefer impulsiveness and remain adaptable.